\documentclass[11pt,twoside]{article}

\usepackage{asp2006}
\usepackage{epsf}
\usepackage{psfig}
\usepackage{lscape}

\begin{document}
\title{Preliminary results for the triple system AV CMi}   
\author{Liakos, A. and Niarchos, P.}   
\affil{Department of Astrophysics, Astronomy and Mechanics,
National and Kapodistrian University of Athens, GR 157 84
Zografos, Athens, Hellas}    

\begin{abstract} 
New CCD photometric observations in V, R and I filters of the
eclipsing binary AV CMi have been obtained. The complete light
curves are analyzed with the Wilson-Devinney code and new
geometric and photometric elements are derived. Moreover, 2-year
systematic observations of the system revealed the existence of a
third body orbiting around one of the components with an
approximate period of 0.52 days. The first light elements for this
additional component are given and its nature is discussed.
\end{abstract}



\section{Introduction}

The system AV CMi (= GSC 07700-0121 = 2MASS J07091084+1211190,
$\alpha_{2000} = 07^{h}09^{m}10.84^{s},~\delta_{2000} =
+12^{\circ}11'19.1''$) has a period of 2.277751 days, an apparent
magnitude of B=11.8 mag and it was discovered by \citet{H68}. The
system was generally neglected, and no complete light curve exists
so far. The only information of its spectral type is given by
\citet{SK90}, while many times of minima have been published. The
secondary component has an eccentric orbit, while the discovery of
a third body with an internal orbit is presented for a first time
in the current study.

\section{Observations and analysis}

In order to obtain a complete light curve (hereafter LC) of the
system, the observations were carried out for 21 nights from
December 2007 to March 2008, while for the tracing of the third
body, systematic observations in I filter took place during the
winters of 2008 and 2009. The system was observed at the Athens
University Observatory, using a 40-cm Cassegrain telescope
equipped with the CCD camera ST-8XMEI and the V, R, I Bessell
photometric filters. Differential magnitudes were obtained by
using the software Muniwin v.1.1.23 \citep{H98}, while the stars
GSC 0770-0929 and GSC 0770-911 were selected as comparison and
check stars, respectively.

The LCs were analysed with the $PHOEBE~0.29d$ software
\citep{PZ05} which uses the 2003 version of the WD code
\citep{WD71,W79}. Due to the lack of spectroscopic mass ratio of
the system, the q-search method was applied in Mode 2, 4 and 5 in
order to find the most probable value of the (photometric) mass
ratio ($q_{ph}$). In each Mode the method of Multiple Subsets was
used in order to obtain the final photometric solution.

The values of the temperatures of the components were adopted
according to the spectral classification of \citet{SK90}, as F2
and G5, respectively. The minimum value of the weighted sum of the
squared residuals was found in Mode 2. The q-search method in this
mode converged to a mass ratio value close to 0.71, which was used
as initial one and then it was adjusted in the subsequent analysis
in the same Mode. The albedos $A_{1},~A_{2}$ and the gravity
darkening coefficients $g_{1},~g_{2}$ were given their theoretical
values according to the spectral type of each component. The
synthetic and observed light curves and the q-search plot of AV
CMi are shown in Fig. 1, the 3-D model of the system in Fig. 2 and
the derived parameters from the light curve solution are listed in
Table 1.

\begin{table}

\caption{The parameters of AV CMi derived from the LCs solution}
\label{tab1} \vspace{1mm}\centering{
 \scalebox{0.95}{
\begin{tabular}{cc|ccccc}

\hline\hline
$Parameter$                 &       $value$     &    $Parameter$     &        \multicolumn{4}{c}{$value$}                           \\
\hline
$q~(m_{2}/m_{1})$           &     0.710~(2)     &                    &      V       &       R       &        I      &               \\
\cline{4-6}
$i$~[deg]                   &       83.8~(4)    &   $x_{1}$$^{**}$   &      0.493   &     0.418     &     0.345     &               \\
$e$                         &       0.11~(1)    &   $x_{2}$$^{**}$   &      0.493   &     0.417     &     0.344     &               \\
$T_1$$^{*}$[K]              &       7000        &   $L_{1}/L_{T}$    &    0.464~(1) &   0.455~(1)   &     0.442~(1) &               \\
$T_2$ [K]                   &      7005~(6)     &   $L_{2}/L_{T}$    &    0.463~(2) &   0.455~(2)   &     0.442~(3) &               \\
$A_1$$^*$=$A_2$$^*$         &         0.5       &   $L_{3}/L_{T}$    &    0.073~(2) &   0.090~(2)   &     0.116~(2) &               \\
\cline{4-7}
$g_1$$^*$=$g_2$$^*$         &         0.32      &                    &    $Pole$    &    $Point$    &     $Side$    &    $Back$     \\
\cline{4-7}
$\Omega_{1}$                &       6.04~(1)    &        $r_1$       &     0.190    &     0.194     &     0.191     &     0.193     \\
$\Omega_{2}$                &       5.29~(1)    &        $r_2$       &     0.176    &     0.180     &     0.177     &     0.179     \\
\hline
$\chi^{2}$                  &       0.21699     &                    &              &               &               &               \\
\hline \hline
\end{tabular}}}
\textit{$^*$assumed}, \textit{$^{**}$\citet{VH93}}, \textit{$L_{T}
= L_{1}+L_{2}+L_{3}$}
\end{table}

\begin{figure}[h!]
\plottwo{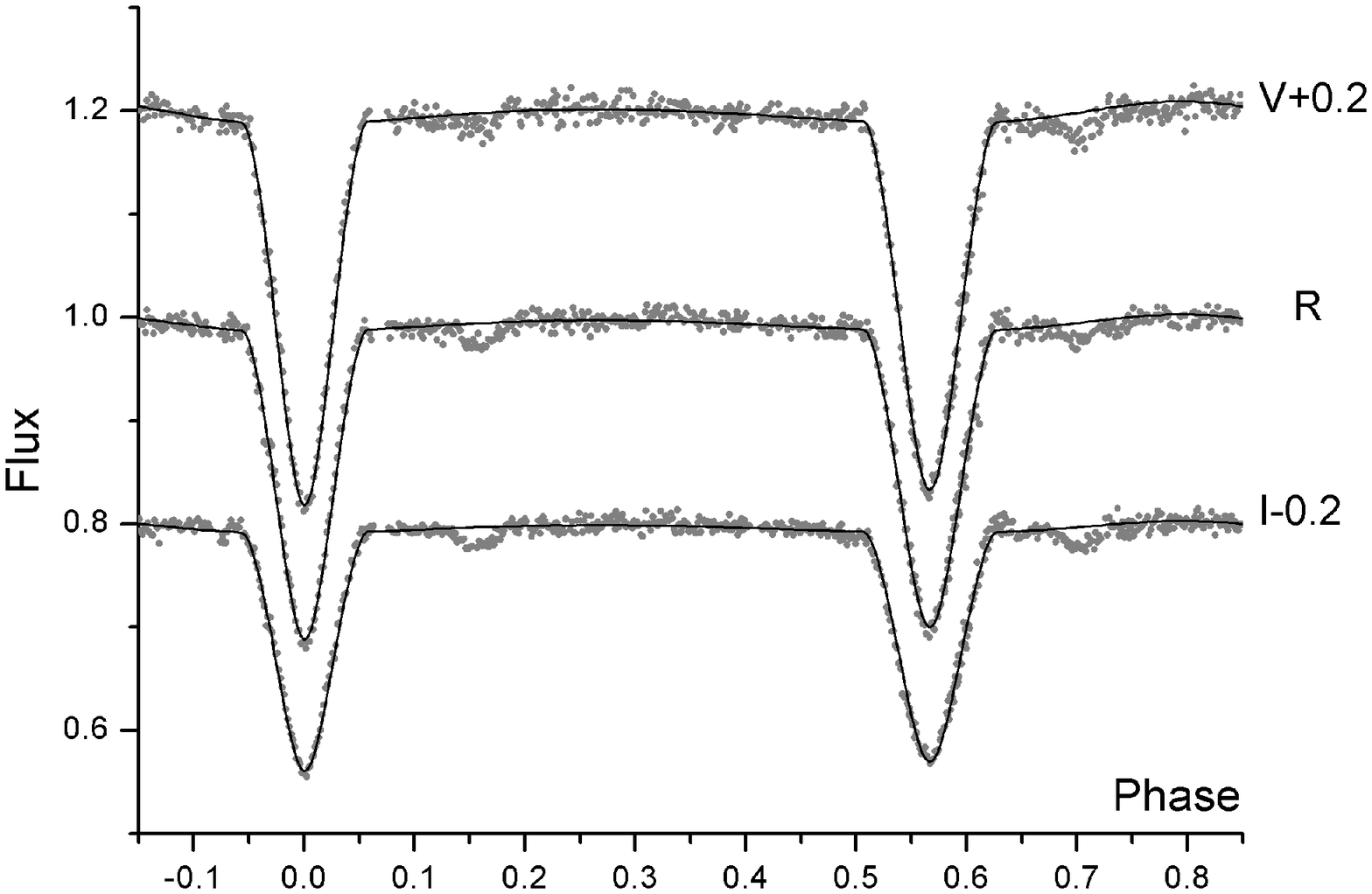}{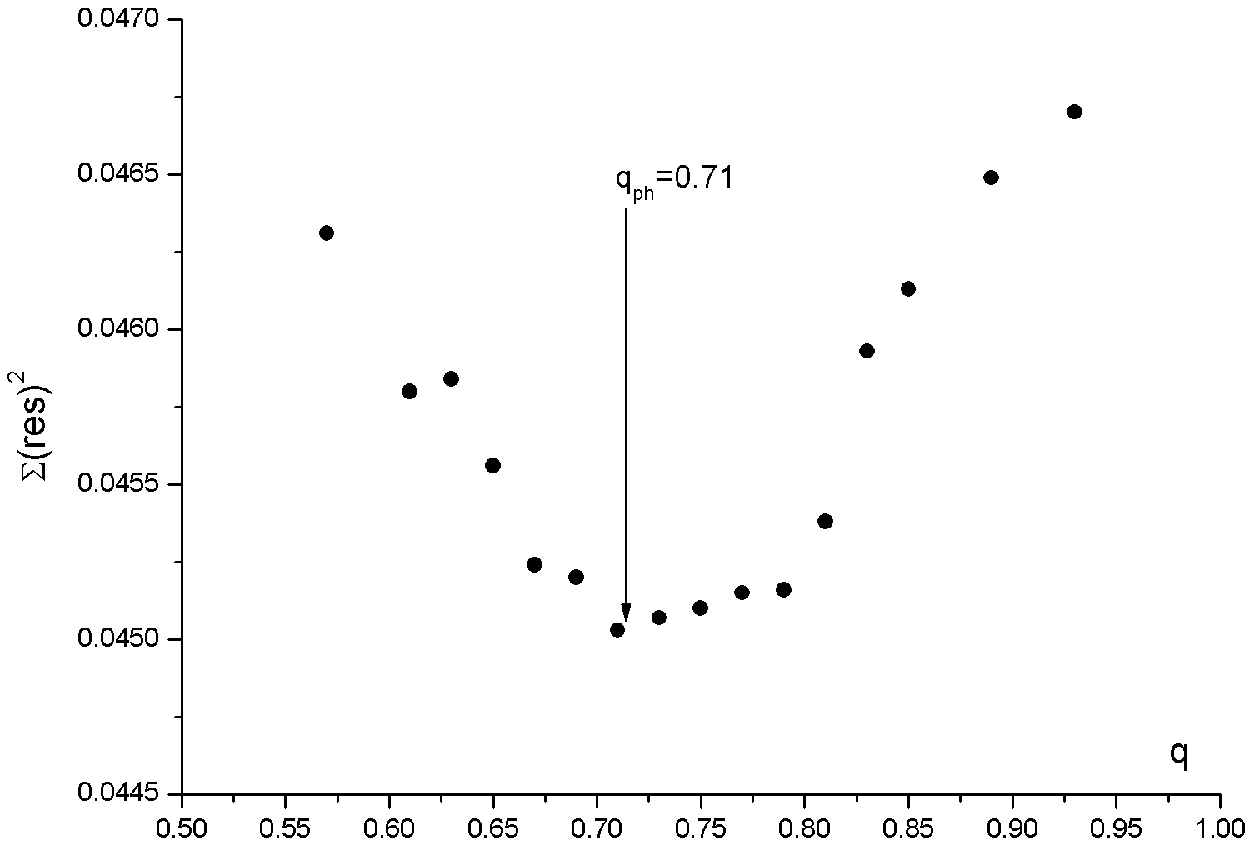}

\caption{Left panel: The synthetic LCs (black solid lines) along
with the observed ones (gray points). Right panel: The q-search
plot of AV CMi in mode 2.}
\end{figure}

\begin{figure}[h!]
\plotone{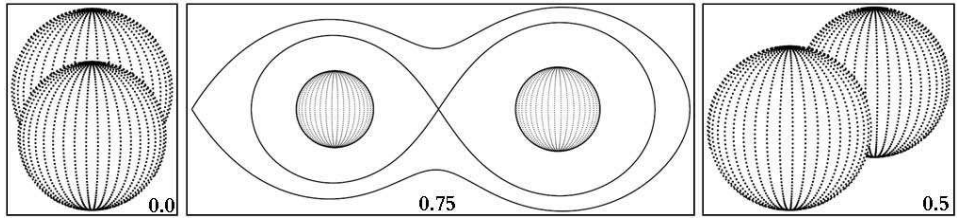}

\caption{The 3-D model of AV CMi in various phases.}
\end{figure}

\section{Light elements of the third body}

From our observations in VRI filters it was found that the maximum
flux drop, during the transit of the third body in front of one of
the components, occurs in I filter. We managed to detect five
transits of the third body (see Fig. 3), to calculate its orbital
period and finally construct its light ephemeris, which is:
$T=HJD~2454521.36315~+~0.5192237^{d}~\times~E$. The duration of
the transit was found to be approximately 3.3 hrs, while the mean
flux drop of the eclipsing binary was found to have a value of
2.6\%.

\begin{figure}[h!]
\plotone{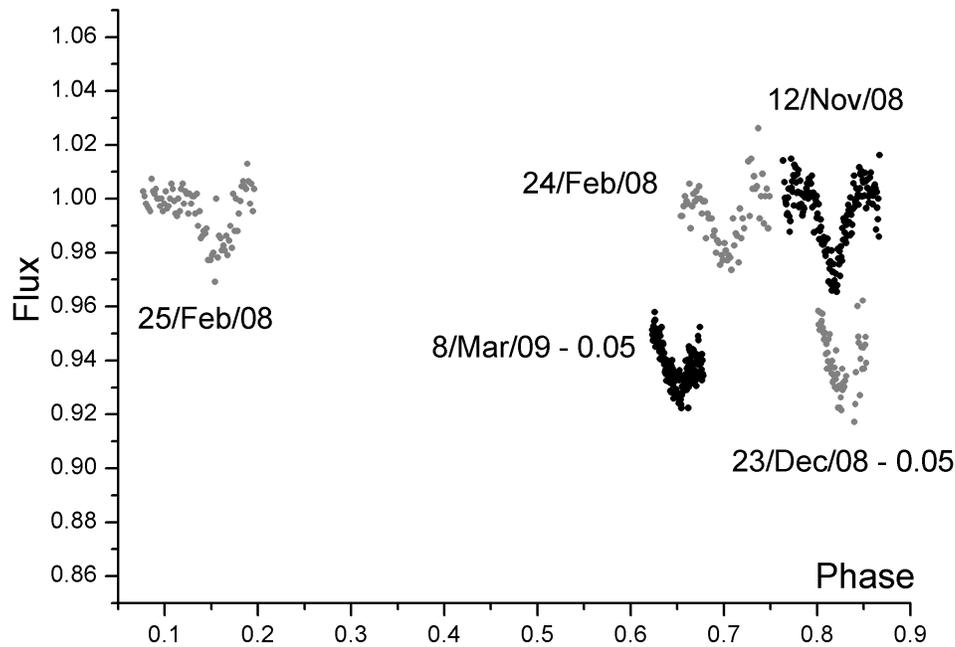}

\caption{The flux drop of AV CMi in I-filter due to the transit of
the third body in front of one of the other components in various
dates.}
\end{figure}

\section{Discussion and Conclusions}

Complete VRI light curves of AV CMi were obtained. The light curve
analysis showed that the system is detached, and its components
have almost the same temperature. The secondary component (less
massive) was found to have eccentric orbit. Moreover, during the
analysis a luminosity contribution of a third body was taken into
account, since the observations revealed the existence of transits
in front of one of the components of the system. The luminosity
fraction of the third body was found to be
$L_3/L_{Total}=9.3(2)\%$, while the maximum luminosity
contribution was found in I-filter, indicating its cool nature. It
has an internal orbit, almost coplanar with the one of the
eclipsing pair, and its period was found to be $\sim$~0.52 days.
It is very difficult to conclude which component is eclipsed by
the third body, since both of them have the same temperature. The
2.6\% flux drop of the eclipsing binary indicates that the size of
the third body is probably very small, and that is the reason we
cannot observe its eclipse, but only its transit. Spectroscopic
observations are certainly needed in order to: i) derive the mass
ratio of the components of the eclipsing pair, ii) define the
spectral types of all the components of the triple system, and
iii) detect periodic shifts in the radial velocity curves due to
the existence of the third body. Moreover, additional photometric
observations in I-filter during the transits will help to
determine the period of the third body with higher accuracy. To
sum up, the preliminary results of the present analysis show that
the third body is small and cool and it has an internal and
eccentric orbit. These characteristics lead us to suggest that the
third body is probably a brown dwarf or a massive Hot Jupiter(!),
but its nature requires further study.

\acknowledgements

This work has been financially supported by the Special Account
for Research Grants No 70/4/9709 of the National \& Kapodistrian
University of Athens, Hellas.




\begin{thebibliography}{}


\bibitem[Hoffmeister(1968)]{H68}
Hoffmeister, C. 1968, AN, 290, 277

\bibitem[Hroch(1998)]{H98}
Hroch, F. 1998, Proc. of the 29th Conference on Variable Star
Research, 30

\bibitem[Pr\v{s}a \& Zwitter(2005)]{PZ05}
Pr\v{s}a, A., \& Zwitter, T. 2005, ApJ, 628, 426

\bibitem[Van Hamme(1993)]{VH93}
van Hamme, W. 1993, AJ, 106, 2096

\bibitem[Svechnikov \& Kuznetsova(1990)]{SK90}
Svechnikov, M.~A., Kuznetsova, E.~F. 1990, Catalogue of
Approximate Photometric and Absolute Elements of Eclipsing
Variable Stars, A.M. Gorky University of the Urals, Sverdlovsk

\bibitem[Wilson \& Devinney(1971)]{WD71}
Wilson, R. E., \& Devinney, E. J. 1971, ApJ, 166, 605

\bibitem[Wilson(1979)]{W79}
Wilson, R. E. 1979, ApJ, 234, 1054
\end{thebibliography}
\end{document}